\documentclass[aps,prb,twocolumn,groupedaddress]{revtex4}
\usepackage{graphicx}
\usepackage{dcolumn}
\begin{document}
\title{Mode identification of high-quality-factor single-defect
    nanocavities \\ in quantum dot-embedded photonic crystals}
\author{Masayuki Shirane}
\email{m-shirane@bu.jp.nec.com}
\author{Shunsuke Kono}
\author{Jun Ushida}
\author{Shunsuke Ohkouchi}
\affiliation{Fundamental and Environmental Research Laboratories,
    NEC Corporation, 34 Miyukigaoka, Tsukuba 305-8501, Japan}
\author{Naoki Ikeda}
\author{Yoshimasa Sugimoto}
\thanks{Present address: National Institute for Material Science, 1-2-1
        Sengen, Tsukuba 305-0047, Japan}
\affiliation{Ultrafast Photonic Devices Laboratory, National
    Institute of Advanced Industrial Science and Technology,
    1-1-1 Umezono, Tsukuba 305-8568, Japan \\
    and Center for Tsukuba Advanced Research Alliance, University of Tsukuba,
    1-1-1, Tennodai, Tsukuba, 305-8577, Japan}
\author{Akihisa Tomita}
\affiliation{Fundamental and Environmental Research Laboratories,
    NEC Corporation \\
    and SORST, Japan Science and Technology Agency,
    34 Miyukigaoka, Tsukuba 305-8501, Japan}
\date{February 9, 2007}
\preprint{\it{Accepted for publication in Journal of Applied Physics}}
\begin{abstract}
We investigate the quality ($Q$) factor and the mode dispersion of
single-defect nanocavities based on a triangular-lattice GaAs
photonic-crystal (PC) membrane, which contain InAs quantum dots
(QDs) as a broadband emitter. To obtain a high $Q$ factor for the
dipole mode, we modulate the radii and positions of the air holes
surrounding the nanocavity while keeping six-fold symmetry. A
maximum $Q$ of 17~000 is experimentally demonstrated with a mode
volume of $V=0.39(\lambda / n)^3$. We obtain a $Q/V$ of
44~000$(n/\lambda)^3$, one of the highest values ever reported with
QD-embedded PC nanocavities. We also observe ten cavity modes within
the first photonic bandgap for the modulated structure. Their
dispersion and polarization properties agree well with the numerical
results.
\end{abstract}
%
%
\maketitle
%

\section{introduction\label{sec:intro}}
Photonic crystal (PC) nanocavities can confine light within a very
small mode volume ($V$) of less than a cubic wavelength with a high
quality ($Q$) factor.\cite{Painter99_H1_laser} %
Because the light-matter interaction is enhanced in such structures,
much attention has been paid to investigating solid-state cavity
quantum electrodynamics in PC nanocavities containing a
semiconductor quantum dot (QD) as a quantum emitter intended for
quantum information processing. Depending on the coupling strength
between a cavity mode and a quantum emitter, physics of the system
can be categorized into either a strong or weak coupling regime.
In the strong coupling regime, the cavity and emitter coherently
exchange energy back and forth (Rabi oscillation),%
\cite{Yoshie04_SC,Reithmaier04_SC,Peter05_SC} %
and this system can be utilized as a qubit.
In the weak coupling regime, the spontaneous emission rate of the
emitter can be enhanced or suppressed compared to the rate obtained
in a vacuum (Purcell effect).%
\cite{Purcell46,Badolato05_S1_QD_UCSB,Gevaux06_H1_QD_Toshiba} %
This can be applied for not only efficient single photon sources%
\cite{Santori02_ind_SPS,Englund05_H1_QD_SPS,Laurent05_H1_QD_SPS_French,
Chang06_L3_QD_SPS} but also polarization-entangled photon
sources based on biexciton-exciton cascade emissions.%
\cite{Benson00_EPR_theory,Stevenson06_EPR,Akopian06_EPR} In the
latter case, the nanocavity enhances the exciton decay rate to
improve the fidelity of the entangled
state,\cite{Stace03_EPR_theory,Troiani06_EPR_cav_theory} and its
cavity mode should be polarization independent.

Now we consider an optimum cavity structure for this purpose. There
are two main types of PC lattice structures: triangular and square.
The first photonic bandgap (PBG) size of the triangular lattice is
much larger than that of the square lattice. A wider PBG is
preferable for a flexible design of high-$Q$ nanocavities. Moreover,
several cavity modes can exist within a wide PBG. In this case, the
double resonance is possible with the proper cavity design, where
one cavity mode is resonant with a ground exciton and another cavity
mode with a first-excited exciton simultaneously. This enables an
effective resonant excitation of the exciton in a
QD.\cite{Nomura06_res_excite2} Several types of high-$Q$ PC
nanocavities based on the triangular lattice have been
experimentally demonstrated, but six-fold symmetry is intentionally
broken to obtain a high $Q$ factor in most of these
cavities.\cite{Akahane03_L3,Song05_Q,Kuramochi06_Q,%
Frederick06_H1_highQ} These cavity modes are not polarization
independent. The six-fold symmetry should be kept to apply cavities
for the polarization-entangled photon
sources.\cite{Hennessy06_AFM_tuning}
\begin{figure*}
\includegraphics[width=170mm]{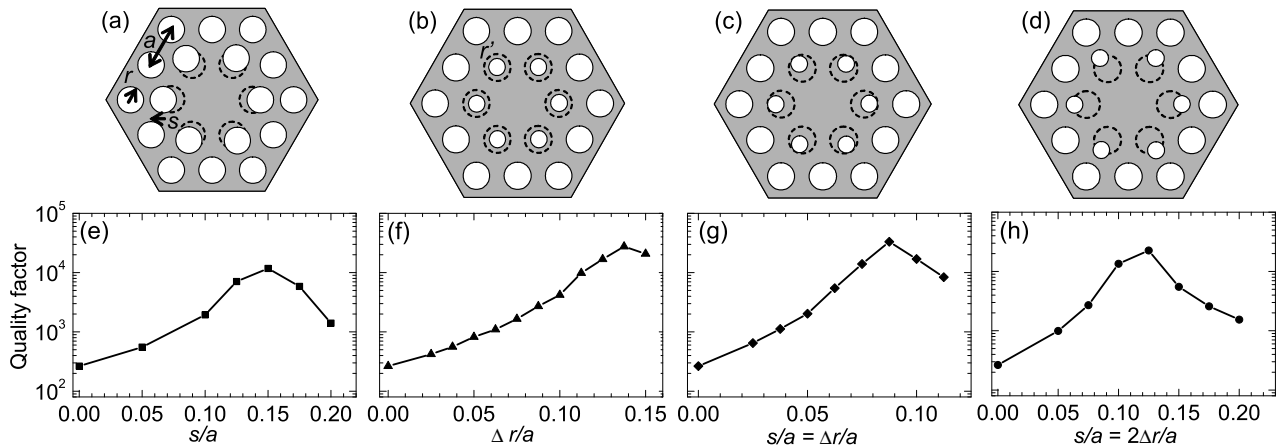}%
    \caption{(a)--(d) Modified H1 nanocavity structures. (e)--(h)
    Calculated $Q$ factor as a function of modulation parameters of
    $\Delta r/a$ and/or $s/a$ for the structures of (a)--(d),
    respectively. The 3D-FDTD method was used with geometrical
    parameters of $r/a=0.31$ and $d/a=0.71$. A maximum $Q$ of
    33~000 was obtained at $\Delta r/a = s/a = 0.09$ as shown in (g).
    \label{fig:qcal}}
\end{figure*}

In this paper, we present our study of the $Q$ factor and the mode
dispersion of single-defect PC membrane nanocavities made in a
triangular lattice of air holes, which contain QDs as a broadband
emitter. To obtain the high $Q$ factor, we modulate the air holes
surrounding the cavity while keeping the six-fold symmetry. In sec.\
\ref{sec:calc}, we present a numerical analysis of the $Q$ factor
and the cavity mode dispersion that we calculated for the modulated
nanocavities. In sec.\ \ref{sec:exp}, the fabrication process of
QD-embedded PC nanocavities and the experimental setup are
explained. In sec.\ \ref{subsec:Q-exp}, we present our measurement
of the $Q$ factor of the dipole mode and comparison with the
calculated one. In sec.\ \ref{subsec:mode-exp}, we present our
investigation of the characteristics of all the cavity modes within
the first PBG. Ten modes are observed at maximum and their
dispersion and polarization properties are compared with the
numerical results. Finally, a brief conclusion is provided.

\section{Numerical analyses\label{sec:calc}}
\subsection{Quality factor of dipole modes%
\label{subsec:Q-calc}}

The nanocavity investigated in this work is a so-called H1
nanocavity based on a triangular lattice of air holes formed in a
thin GaAs membrane with one hole removed.\cite{Painter99_H1_theory}
To obtain a high $Q$ factor for the dipole mode, we modulated the
radii and positions of the six nearest-neighbor holes to the cavity
while keeping the six-fold symmetry.%
\cite{Huh02_H1_mono_mode,Park02_H1_mode,Ryu03_H1_hexapole,Dalacu06_H1_mode}
Though several ways can be used to perform the modulation, we simply
changed the radii of the six air holes and/or their positions.
Figures \ref{fig:qcal}(a)--(d) show schematic diagrams of four types
of modulated H1 nanocavities. In these figures, $a$ is a lattice
constant and $r$ ($r'$) is a regular (modulated) hole radius. In
Fig.\ \ref{fig:qcal}(a), the six holes are shifted from the original
positions (shown with dashed lines) to outside the cavity by $s$
along the lines of symmetry with those radii fixed to $r' = r$. In
Fig.\ \ref{fig:qcal}(b), the radii of the six holes are reduced by
$\Delta r = r-r'$ without any position shifts. In Figs.\
\ref{fig:qcal}(c) and (d), both the radii and position of the six
holes are modified under the conditions of $s = \Delta r$ for (c)
and $s = 2 \Delta r$ for (d).

The three-dimensional (3D) finite-difference time-domain (FDTD)
method was used to calculate the $Q$ factors.\cite{Yee66_FDTD} The
parameters included $r = 0.31a$, a slab thickness of $d = 0.71a$,
and a refractive index of $n=3.4$ for GaAs. The size of the
calculation domain was set to $31 a \times 16 \sqrt{3} a$ in the
slab plane and to $6a$ in the vertical direction to the slab, and
the cavity was put at its center. The grid size was set to $a/16$.
The calculation domain was surrounded by the perfect matched layer
(PML) based on Mur's second-order absorbing interface condition.
The total energy stored in the cavity $U(t)$ is expressed as
follows:
\begin{equation}\label{eq:Q}
    U(t) = U_0 \exp \left( -\frac{\omega_0}{Q} t \right),
\end{equation}
where $\omega_0$ is the angular frequency of the cavity mode derived
from the Fourier transform of the electric field, $E(t)$. As can be
seen in Eq.\ (\ref{eq:Q}), the $Q$ factor can be estimated from the
slope of the exponential decay of the total energy within the
cavity.
Figures \ref{fig:qcal}(e)--(h) show calculated $Q$ factors as
functions of modulation parameters corresponding to Figs.\
\ref{fig:qcal}(a)--(d). As $r'$ is reduced and/or $s$ is increased,
the $Q$ factor increases drastically and then decreases. A maximum
$Q$ of 33~000 was obtained at $\Delta r/a = s/a = 0.09 $ as shown in
Fig.\ \ref{fig:qcal}(g). The mode volume was estimated to be $V =
0.39 (\lambda/n)^3$ for the maximum-$Q$ structure.

\subsection{Cavity mode dispersion and field distribution
\label{subsec:modes-calc}}
As mentioned earlier, the six holes were reduced and shifted to
outside the cavity to obtain a high $Q$ factor for the dipole mode.
Such modulations slightly enlarge the cavity size, and accordingly,
the energy of the cavity modes decreases.
\begin{figure}
\includegraphics[width=85mm]{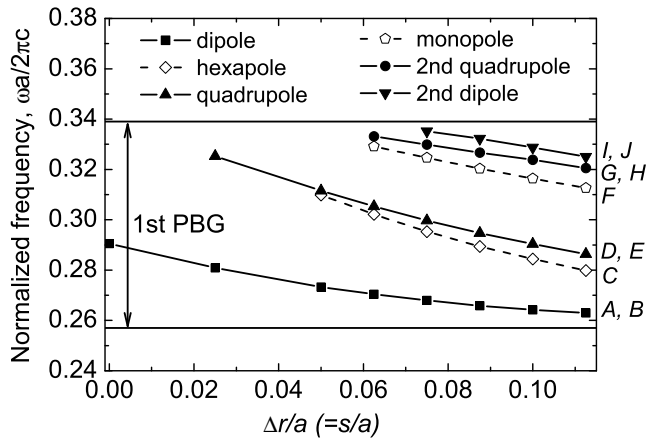}%
    \caption{Normalized cavity-mode frequencies as a function of
    $\Delta r/a$ ($= s/a$) calculated using the 3D-FDTD method.
    The solid (dashed) lines correspond to doubly degenerate
    (non-degenerate) cavity modes. The ten modes are referred to as
    modes \textit{A--J} in order of frequency.
    \label{fig:modecal}}
\end{figure}
Figure \ref{fig:modecal} shows the calculated cavity-mode
frequencies as a function of $\Delta r/a$ ($= s/a$) [see Fig.\
\ref{fig:qcal}(c)] within the first PBG. The 3D-FDTD method was used
again including geometrical parameters of $r/a=0.31$ and $d/a=0.71$.
The solid (dashed) lines correspond to doubly degenerate
(non-degenerate) cavity modes. Without modulation ($\Delta r/a =
0$), only the degenerate dipole modes exist within the PBG. As the
radii decrease, several modes that originally exist in the second
band without modulation appear within the PBG and move toward the
first band. Ten modes including the degeneracy exist within the PBG
at $\Delta r/a \geq 0.075$. A similar mode dispersion was reported
by Park \textit{et al}.\ theoretically.\cite{Park02_H1_mode} With
the FDTD method, the energy of the cavity modes can be estimated
from the Fourier transform of $E(t)$. The degeneracy of the modes
can be checked by introducing parity conditions. The plane-wave
expansion (PWE) method was also used to reconfirm it.
\cite{Ho_90_PWEM,Johnson01_PWEM}

\begin{figure}
\includegraphics[width=85mm]{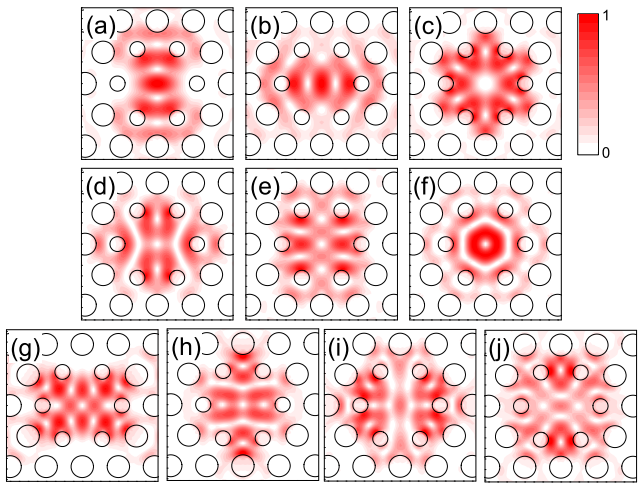}%
    \caption{Calculated electric field distributions
    [$|E|/\textrm{max}(|E|)$] in the slab center for ten cavity
    modes with $\Delta r / a = 0.11$.
    The modes in (a)--(j) correspond to modes \textit{A--J} in
    Fig.\ \ref{fig:modecal}, respectively. \label{fig:modefld}}
\end{figure}
Figure \ref{fig:modefld} shows the calculated electric field
distributions [$|E|$/max($|E|$)] in the slab center for ten cavity
modes with $\Delta r / a = 0.11$. Based on their mode shapes, they
are called degenerate dipole [Fig.\ \ref{fig:modefld}(a) and (b)],
hexapole [Fig.\ \ref{fig:modefld}(c)], degenerate quadrupole [Fig.\
\ref{fig:modefld}(d) and (e)], monopole [Fig.\
\ref{fig:modefld}(f)], degenerate second-quadrupole [Fig.\
\ref{fig:modefld}(g) and (h)], and degenerate second-dipole modes
[Fig.\ \ref{fig:modefld}(i) and (j)], respectively, from the bottom
to the top in the band diagram. For simplicity, the modes in Fig.\
\ref{fig:modefld}(a)--(j) are called modes \textit{A--J}. In the
real case, however, the degeneracy breaks due to small local
fabrication error, which will be discussed in detail later. To
create a strong interaction between a cavity mode and a single QD,
we should locate the QD at the antinode of the field. Since only a
few groups can use a site-controlled QD,\cite{Badolato05_S1_QD_UCSB}
most use self-assembled or monolayer-fluctuated QDs whose positions
cannot be controlled. Once a site-control technique is available,
cavity modes that have a low number of nodes are desirable because
they can withstand a position error better than those with a lot of
nodes. From this point of view, modes $A$, $B$, and $F$ are
preferable.

\begin{figure}
\includegraphics[width=75mm]{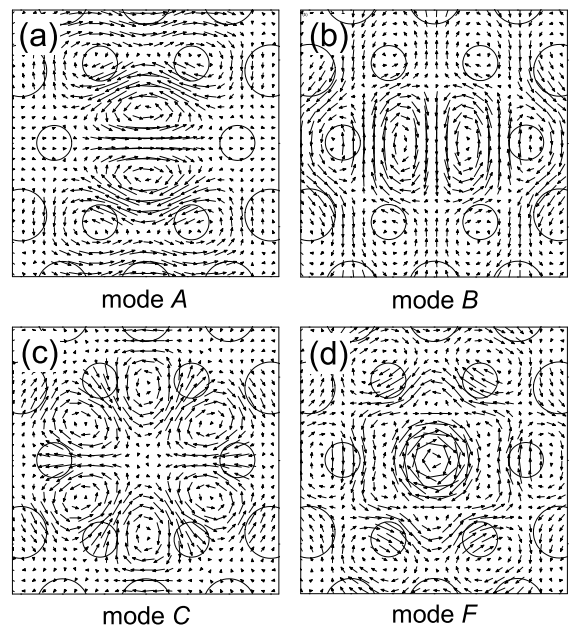}%
    \caption{Calculated electric field directions of (a) mode $A$, (b) $B$,
    (c) $C$, and (d) $F$. Modes $A$ and $B$ can be utilized for
    manipulating polarization-entangled states. Although
    non-degenerate modes of $C$ and $F$ have no preferred direction
    in the far fields, they cannot be applied for this purpose.
    \label{fig:vecfld}}
\end{figure}
Figure \ref{fig:vecfld}(a) and (b) show the electric field
directions of modes $A$ and $B$, respectively. These modes are
ideally degenerate, and both modes have a strong electric field at
the center of the cavity. Not only the far field directly above the
cavity but also the local field at the center has no preferred
direction. A single QD should be located at this position to
manipulate polarization-entangled states. On the other hand, mode
$F$ is non-degenerate. This mode radially oscillates in the
slab-plane with respect to the cavity center. Although the far field
directly above the cavity is unpolarized, the local field at the
antinode has a preferred direction, as shown in Fig.\
\ref{fig:vecfld}(d). Therefore, mode $F$ can be applied for single
photon sources and a strong coupling system but not for
polarization-entangled photon sources. Mode $C$, whose electric
field direction is shown in Fig.\ \ref{fig:vecfld}(c), is also
non-degenerate and the situation is the same as with mode $F$.

Four modes are at the slightly higher energy side of mode $F$. For
example, for $a = 300$ nm, the energy (wavelength) of mode $F$ is
1.336 eV (924 nm), and the energies of mode $G$ and $I$ are 34 and
53 meV larger than that of mode $F$. These values are similar to
typical energy separations between the ground and first-excited
excitons in self-assembled InAs/GaAs quantum
dots.\cite{Santori04_QD_correlation} Modes $F$ and $G$ (or $I$) may
be simultaneously resonant with the ground and first-excited
excitons, respectively. When a pump light is tuned to the
first-excited exciton, it is enhanced by the latter cavity mode and
an electron-hole pair is effectively created only in the specific
QD.\cite{Nomura06_res_excite2} This resonant excitation suppresses
the generation of excess carriers that degrade the quality of single
and entangled photon sources. The ground exciton also interacts with
the former cavity mode, which results in a Purcell effect or strong
coupling.
%

\section{Experiments \label{sec:exp}}
\begin{figure}
\includegraphics[width=85mm]{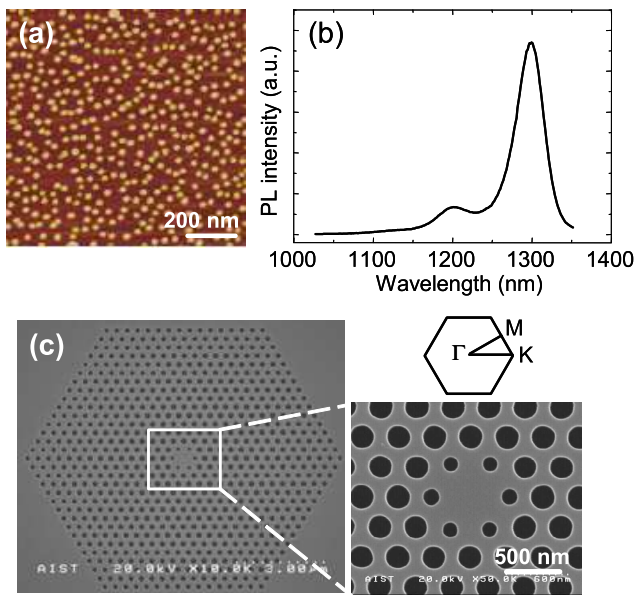}%
    \caption{Specifications of fabricated QD-embedded PC nanocavity.
    (a) AFM image of the surface of a QD layer. (b) PL spectrum of the
    ensemble QDs at room temperature. (c) SEM image of the PC
    structure. The corresponding reciprocal lattice space is also
    shown. \label{fig:fab}}
\end{figure}
Based on the calculation results, we fabricated H1 nanocavities with
the modulation shown in Fig.\ \ref{fig:qcal}(c). They contain the QD
ensemble which functions as a broadband emitter to probe the cavity
modes. The samples were grown by molecular-beam epitaxy on a (001)
GaAs substrate. A 248-nm-thick GaAs membrane was grown on a
2-$\mu$m-thick Al$_{0.6}$Ga$_{0.4}$As sacrificial layer on the
substrate. A single layer of InAs self-assembled QDs were embedded
in the middle of the GaAs membrane layer. The strain-relaxation
layer was made on the QD layer to obtain longer wavelength
emissions. After the growth of the membrane, H1 nanocavities were
made using electron-beam (EB) lithography, dry etching, and
selective wet-etching techniques.
\cite{Sugi04_fab} %
A reference QD sample grown with the same condition but without
layers above and air holes was also prepared. Figure
\ref{fig:fab}(a) shows an atomic force microscope (AFM) image of the
surface of the reference sample. The density of the QDs was
estimated to be about $3 \times 10^{10}$ cm$^{-2}$ from the AFM
image. The average height and diameter of the QDs were also
estimated to be 4 and 30 nm, respectively. Figure \ref{fig:fab}(b)
shows the photoluminescence (PL) spectrum of the fabricated QD
ensemble at room temperature. We see not only the first peak
($\lambda$ $\sim$1300 nm), which corresponds to the ground exciton
transitions, but also the second peak ($\lambda$ $\sim$1200 nm), which
corresponds to the first-excited exciton transitions. Figure
\ref{fig:fab}(c) shows a scanning electron microscope (SEM) image of
a typical H1 nanocavity. A number of nanocavities were prepared by
changing structural parameters, a regular air hole radius, $r$, and
a modulated air hole radius, $r'$, while the lattice constant was
fixed to $a = 350$ nm. The H1 nanocavity was surrounded by 15
periods of air holes for good in-plane optical confinement.

The samples were set in a conduction-type microscope cryostat that
cooled them down to 4.3 K. The cryostat was mounted on a
piezo-actuated translation stage for precise sample
positioning.\cite{Kono05_exp} The samples were optically excited by
a He-Ne laser ($\lambda = 633$ nm). The pump beam was focused by a
microscope objective lens (numerical aperture of 0.42) to a spot
size of 4 $\mu$m on the sample. The pump power was set to $\sim$3
kW/cm$^2$. The PL was collected by the same objective lens and
spectrally resolved by a single-grating monochromator (focal length
of 640 mm) equipped with a liquid nitrogen-cooled InGaAs
multichannel detector (MCD, Jobin-Yvon IGA-512). The wavelength
resolution of the system was 35 pm at $\lambda = 1310$ nm with a
1200-groove/mm grating. This was limited by the 50-$\mu$m pixel
width of the MCD. To obtain a higher wavelength resolution, we used
a liquid nitrogen-cooled photomultiplier (PMT, Hamamatsu Photonics
R5509-42) instead of the MCD, and both the entrance and exit slit
widths of the monochromator were set to 20 $\mu$m. In the latter
case, the wavelength resolution of the system was 19 pm.
%

\section{Results and discussion \label{sec:results}}
\subsection{Quality factor measurement of dipole modes%
\label{subsec:Q-exp}}%
\begin{figure}
\includegraphics[width=85mm]{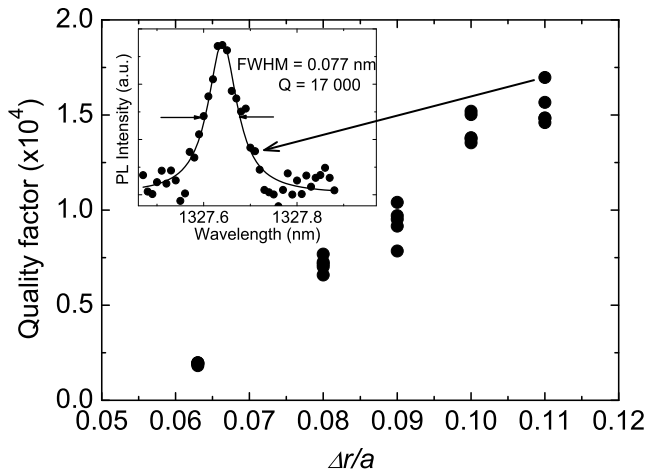}%
    \caption{Measured $Q$ factor as a function of the modulation
    parameter of $\Delta r/a$. A maximum $Q$ of 17~000 was obtained
    at $\Delta r/a = 0.11$. The inset shows a spectrum of the
    maximum-$Q$ nanocavity measured with the PMT. \label{fig:qexp}}
\end{figure}
The $Q$ factor of the dipole modes was extracted from the Lorenzian
fit to the PL peaks measured at room temperature. Because the PL
spectrum is the Fourier transform of Eq.\ (\ref{eq:Q}), $Q$ is equal
to $\lambda / \Delta \lambda$, where $\Delta \lambda$ is the full
width at half maximum (FWHM) of the PL spectrum. Figure
\ref{fig:qexp} shows the obtained $Q$ factor as a function of the
modulation parameter. The regular air-hole radius was $r=114$ nm
($r/a = 0.33$). The modulation parameters of $r'$ and $s$ were
estimated from the SEM image of the nanocavities. The PL-peak
wavelengths were $\sim$1300 nm, which correspond to the PL-center
wavelengths of the QD ensemble. Five nominally identical cavities
for each modulation parameter were investigated. As $\Delta r/a$
increases, the $Q$ factor rapidly increases and it becomes more than
13~000 at $\Delta r/a$ = 0.10--0.11. The change in $Q$ factor agrees
well with the calculated one shown in Fig.\ \ref{fig:qcal}(g). The
cavity wavelength redshifts as $\Delta r/a$ increases because the
effective cavity size increases. This change is plotted in Fig.\
\ref{fig:modeexp}. The maximum $Q$ of 17~000 was obtained for one of
the cavities with $\Delta r/a$ = 0.11. The PL spectrum for this
cavity taken with the PMT is shown in the inset of Fig.\
\ref{fig:qexp}. The calculated mode volume, which is much less
sensitive to the modulation parameter compared to the $Q$ factor, is
$V = 0.39(\lambda/n)^3$ for this cavity. The obtained $Q/V$, which
is the figure of merit in the weak coupling regime, is
44~000$(n/\lambda)^3$. This is one of the
highest values ever reported with QD-embedded PC nanocavities.%
\cite{Yoshie04_SC,Badolato05_S1_QD_UCSB,Frederick06_H1_highQ,%
Hennessy06_AFM_tuning} The expected Purcell factor is $F_P = 3Q
(\lambda/n)^3 / 4 \pi^3 V = {}$3300 when a QD is located at the
cavity center and when the cavity mode is right resonant with the QD
exciton. The measured maximum-$Q$ is lower than the calculated $Q$
of 33~000 by factor of 2, and we think that the measured $Q$ is
limited by the QD absorption and the fabrication error through the
EB lithography and dry etching process.

\begin{figure}
\includegraphics[width=80mm]{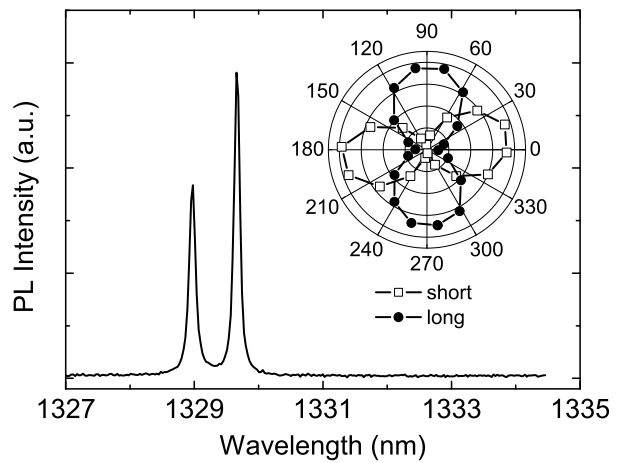}%
    \caption{Typical spectrum of the dipole modes with $Q>$ 10~000.
    The small local fabrication error breaks the degeneracy of modes
    $A$ and $B$ to make a 0.7-nm split. Inset: polar plot of
    the polarization dependence of short (open squares with line)
    and long (filled circles with line) wavelength modes.
    The geometry is the same as that shown in Fig.\
    \ref{fig:fab}(c). \label{fig:poldep}}
\end{figure}
Figure \ref{fig:poldep} shows a typical spectrum of dipole modes
with $Q>$ 10~000. The dipole modes $A$ and $B$ should be ideally
degenerate. However, the degeneracy breaks due to the local
imperfection in the fabricated
structure.\cite{Painter02_H1_poldep,Hennessy06_AFM_tuning} The inset
in Fig.\ \ref{fig:poldep} shows the polarization dependence of short
(open squares with line) and long (filled circles with line)
wavelength modes. Its geometry is the same as that shown in Fig.\
\ref{fig:fab}(c). They were linearly polarized: one is parallel to
the $\Gamma$-K direction and the other to the $\Gamma$-M direction.
The typical split was $\delta=$ 1--2 nm. To analyze the effect of
the small local fabrication error, we carried out an additional
modulation to one of the six nearest-neighbor holes in the FDTD
simulation. When the radius of the hole was reduced by 1 nm, the
cavity modes split by 0.7 nm ($\delta / \lambda = 0.05$\%).
Controlling the hole size with less than 1-nm error is difficult
even with a state-of-the-art fabrication technique. Therefore, the
degeneracy can be kept for low-$Q$ ($<1000$) cavities, but the mode
split is unavoidable for high-$Q$ ($>$10~000) cavities. A
polarization-dependent cavity-frequency trimming technique, such as
surface nano-oxidization by the AFM,\cite{Hennessy06_AFM_tuning} is
required to compensate for it.
%

\subsection{Cavity mode identification within the photonic bandgap%
\label{subsec:mode-exp}}%
Figure \ref{fig:modecal} shows that ten cavity modes theoretically
exist within the first PBG at maximum. The mode identification of
the fabricated structures was done by the PL measurement at 4 K. The
wavelengths of the cavity modes and QD ensemble emissions were
shifted to shorter wavelengths by about 20 and 90 nm compared to
those at room temperature, respectively.

\begin{figure*}
\includegraphics[width=148mm]{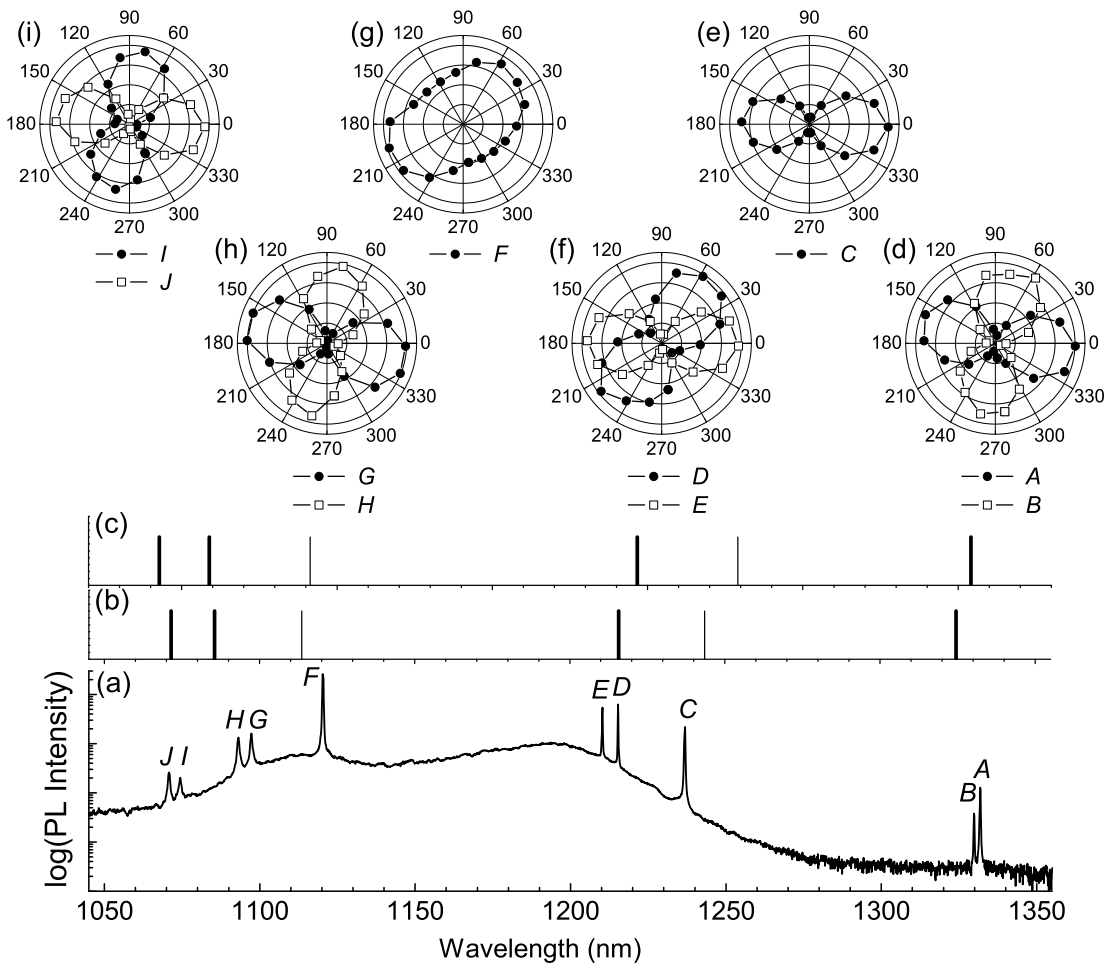}%
    \caption{(a) Wide-range PL spectrum of high-$Q$ nanocavity with
    $\Delta r /a =0.11$ measured at 4 K. (b) and (c) Calculated
    spectra obtained using the FDTD and PWE methods respectively,
    where bold (thin) lines represent degenerate (non-degenerate)
    modes.
    The horizontal axis in (c) is shifted by $-25$ nm compared to the
    others in (a) and (b).
    (d)--(i) Polar plots of polarization dependence of modes
    \textit{A--J} in (a). \label{fig:widespe}}
\end{figure*}
Figure \ref{fig:widespe}(a) shows a wide-range PL spectrum of the
high-$Q$ sample with $r/a=0.31$ and $\Delta r /a$ = 0.11. The
wavelength resolution of this spectrum is 0.1 nm. Ten sharp peaks
were observed in broadband emissions of the QD ensemble. Though
modes $A$ and $B$ are far detuned from the PL center of the QD
ensemble, other emitters such as impurities are likely to contribute
to the emissions. The calculated spectra obtained using the FDTD and
PWE methods are shown in Fig.\ \ref{fig:widespe}(b) and (c),
respectively. In these figures, bold (thin) lines represent doubly
degenerate (non-degenerate) modes. The measured wavelengths agree
well with the calculated ones. The horizontal axis in Fig.\
\ref{fig:widespe}(c) is shifted by $-25$ nm to enable the results to
be compared easily.\cite{footnote1} This is the first demonstration
wherein ten cavity modes were clearly obtained for a single PC
nanocavity. Conventionally, mode dispersion has been measured by
plotting normalized frequencies taken from several samples having
different geometrical parameters. Dalacu \textit{et al}.\ observed
six cavity modes for an H1 nanocavity, however, they did not observe
modes \textit{G--J}.\cite{Dalacu06_H1_mode} The measured split for
the \textit{ideally} degenerate modes were 1.7 nm (modes $A$ and
$B$), 5.1 nm ($D$ and $E$), 4.2 nm ($G$ and $H$), and 3.9 nm ($I$
and $J$). Because the positions of the antinodes in the higher modes
are close to the modulated air holes [See Fig.\ \ref{fig:modefld}],
the local fabrication error affects the higher modes more than the
fundamental modes. This resulted in a larger split of the higher
modes compared to the fundamental modes of $A$ and $B$.

Figures \ref{fig:widespe}(d)--(i) show the polarization dependence
of the ten modes. In Fig.\ \ref{fig:widespe}(g), the monopole mode
$F$ does not have a clear preferred polarization direction as was
theoretically predicted. The non-circular shape of the data is
mainly attributed to the polarization-dependent optical components.
The hexapole mode $C$ is linearly polarized along the $\Gamma$--K
direction as shown in Fig.\ \ref{fig:widespe}(e), though this mode
should be unpolarized theoretically. This can be explained as
follows: As can be seen in Fig.\ \ref{fig:vecfld}(c), this mode
consists of three identical components that oscillate along the
$\Gamma$--K direction. The three components combined equally, and
the mode does not have a preferred direction when six-fold symmetry
is held. Once the symmetry breaks, one of the three components is
dominant, and the mode is polarized parallel to one of the
$\Gamma$--K directions. Modes $A$ and $B$, $G$ and $H$, and $I$ and
$J$ are pairs of orthogonal polarization [Figs.\
\ref{fig:widespe}(d), (h), and (i)], respectively, whereas modes $D$
and $E$ are not orthogonal but their polarization directions differ
by $\sim$45$^\circ$ [Fig.\ \ref{fig:widespe}(f)].\cite{footnote2}
This non-orthogonal polarization property was observed in our
similar structures and elsewhere.\cite{Dalacu06_H1_mode}

\begin{figure}[t]
\includegraphics[width=80mm]{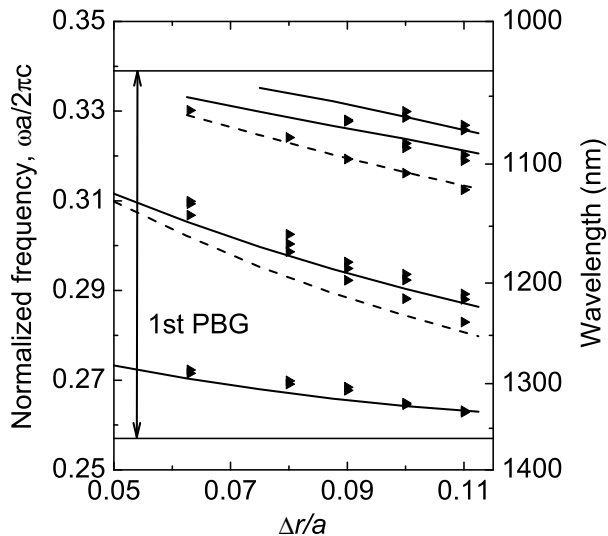}%
    \caption{Measured cavity mode wavelengths (filled triangles) as a
    function of the modulation parameter. The calculated ones are also
    shown with solid (degenerate) and dashed (non-degenerate) lines
    (See Fig.\ \ref{fig:modecal}).
    \label{fig:modeexp}}
\end{figure}
Figure \ref{fig:modeexp} shows the measured cavity mode wavelengths
as a function of the modulation parameter. The calculated
wavelengths are also shown with solid and dashed lines. The measured
ones agree well with the theoretical ones. Though the simulation
predicted that ten cavity modes exist at $\Delta r/a \geq 0.075$, we
experimentally observed ten cavity modes only at $\Delta r/a \geq
0.10$. This is mainly because the bandedge frequency of the second
band of the fabricated structure is lower than the calculated one.
This also affects the $Q$ factor of modes \textit{G--J}. Because the
cavity mode is closer to the bandedge, the in-plane optical
confinement decreases, as does the $Q$ factor. The measured $Q$ of
modes \textit{G--J} was $\sim$1500 at $\Delta r/a=0.11$, which is
more than 10 times lower than the calculated one.
%

\section{Conclusions\label{sec:concl}}
We have theoretically and experimentally investigated the $Q$ factor
and mode dispersion of QD-embedded H1 nanocavities. By modulating
the six air holes surrounding the nanocavity while keeping six-fold
symmetry, we experimentally obtained a maximum $Q$ of 17~000 for the
dipole mode with a very small $V=0.39(\lambda/n)^3$. These values
correspond to a Purcell factor of $F_P =3300$ under an optimum
condition. The measured $Q$ was compatible with a calculated $Q$ of
33~000. Though the dipole mode was designed to be doubly degenerate,
the small fabrication imperfection slightly broke the symmetry,
which caused a mode split of about 0.1\%. Ten cavity modes were
observed in the single H1 nanocavity within the first PBG, and their
dispersion and polarization properties agreed well with those
simulated. The measured energy difference between some higher modes
was almost the same as that between the ground and first-excited
excitons in a QD, which would meet the doubly resonant condition.
Once the cavity modes are precisely tuned to the exciton levels in a
QD, efficient non-classical photon sources can be achieved for
quantum information processing.

\bibliographystyle{apsrev}

\begin{thebibliography}{36}
\bibitem{footnote1}
Although the absolute value of the calculated wavelengths obtained
using the PWE method was about 2\% larger than that obtained using
the FDTD method, the relative wavelength difference between the
modes was quite similar. Therefore, we conclude that both
calculations produce the same results.
\end{thebibliography}

\end{document}